\documentclass[a4paper,11pt]{article}
\usepackage{pos}
\title{Open heavy-flavour production from small to large collision systems with ALICE at the LHC}
\ShortTitle{Open heavy-flavour production with ALICE}

\manuallySeparateAuthors
\author*{Luuk Vermunt}
\author{on behalf of the ALICE Collaboration}
\affiliation{Institute for Subatomic Physics, Utrecht University}
\emailAdd{luuk.vermunt@cern.ch}

\abstract{
Heavy quarks are effective probes to study the properties of the quark--gluon plasma (QGP) produced in ultra-relativistic heavy-ion collisions. In this contribution, the latest results on open heavy-flavour production in pp and Pb--Pb collisions measured by the ALICE Collaboration will be discussed. Measurements of open heavy-flavour production in Pb--Pb collisions allow for testing the mechanisms of heavy-quark transport and energy loss in the medium. In addition, measurements of the elliptic ($v_{2}$) and triangular ($v_{3}$) flow coefficients of heavy-flavour particles give insights into the participation of heavy quarks in the collective motion of the medium, the path-length dependence of their in-medium energy loss, and recombination effects during the hadronisation. Finally, the directed flow ($v_{1}$) of open heavy-flavour particles is sensitive to the unprecedentedly strong magnetic fields present in the early stages of the collision, and so measurements of its charge dependence are key to constraining the electrical conductivity of the QGP. In small hadronic systems like pp, open heavy-flavour production provides the baseline for the investigation of hot-medium effects in heavy-ion collisions, as well as tests of perturbative quantum chromodynamic calculations.
}

\FullConference{
  40th International Conference on High Energy physics - ICHEP2020\\
  July 28 - August 6, 2020\\
  Prague, Czech Republic (virtual meeting)
}

\begin{document}
\maketitle

\section{Introduction}
Charm and beauty quarks are predominantly produced in the early stage of the collisions via hard-scattering processes and are therefore described with perturbative quantum chromodynamic (pQCD) calculations. Hence, measurements of open heavy-flavour hadron production in pp collisions test these pQCD model predictions. By comparing the production of different heavy-flavour hadron species, the hadronisation of heavy quarks can be investigated as well. The latter is particularly interesting for heavy-flavour baryons, since it has been observed that heavy-quark hadronisation into baryons is not well understood in pp collisions~\cite{Acharya:2017kfy, Aaij:2019pqz, Chatrchyan:2012xg}. 

In ultra-relativistic heavy-ion collisions, a phase transition of nuclear matter to a colour-deconfined medium is predicted, the so-called quark--gluon plasma (QGP). Due to the very short time scales characterising open heavy-flavour production, which are shorter than the QGP formation time, heavy quarks experience the full evolution of the medium. Once produced, these probes traverse the medium and interact via inelastic and elastic processes with its constituents. They are therefore an effective probe to study several aspects of the medium, like the microscopic nature of the energy loss mechanisms, the relevance of quark recombination in the medium, and the initial conditions of the system. Detailed reviews of heavy-flavour physics in both small and large collision systems can be found in Refs.~\cite{Andronic:2015wma, Prino:2016cni}.

Open charmed hadrons are measured by ALICE at midrapidity ($|y|<0.5$) via the decay channels ${\rm D}^0  \to {\rm K}^- \pi^+$, ${\rm D}^+ \to {\rm K}^- \pi^+ \pi^+$, ${\rm D}^{*+} \to {\rm D}^0 \pi^+$, $\rm D_{s}^+ \to \phi \pi^+ \to K^{+}K^{-} \pi^+$, ${\rm \Lambda_{c}^+ \to pK^{0}_{s} \to p\pi^+\pi^-}$, ${\rm \Lambda_{c}^{+}\to p K^-\pi^+}$, and their charge conjugates. The charmed-hadron raw yields are extracted via an invariant-mass analysis after having applied topological and particle-identification selections to enhance the signal-over-background ratio. The reconstruction efficiencies are estimated using Monte Carlo simulations and the fraction of prompt charmed hadrons is calculated using a \textsc{FONLL}-based approach~\cite{Cacciari:1998it,Acharya:2019mgn}. In addition, electrons coming from semileptonic decays of heavy-flavour hadrons are measured by ALICE. The electrons are identified at midrapidity using particle identification in the relevant central-barrel detectors~\cite{Aamodt:2008zz}, after which the hadron contamination and electrons from non-heavy-flavour sources are subtracted from the measured inclusive yield~\cite{Acharya:2019mom}.

\section{Results}
\begin{figure}[!tb]
	\begin{center}
		\includegraphics[width=.495\textwidth]{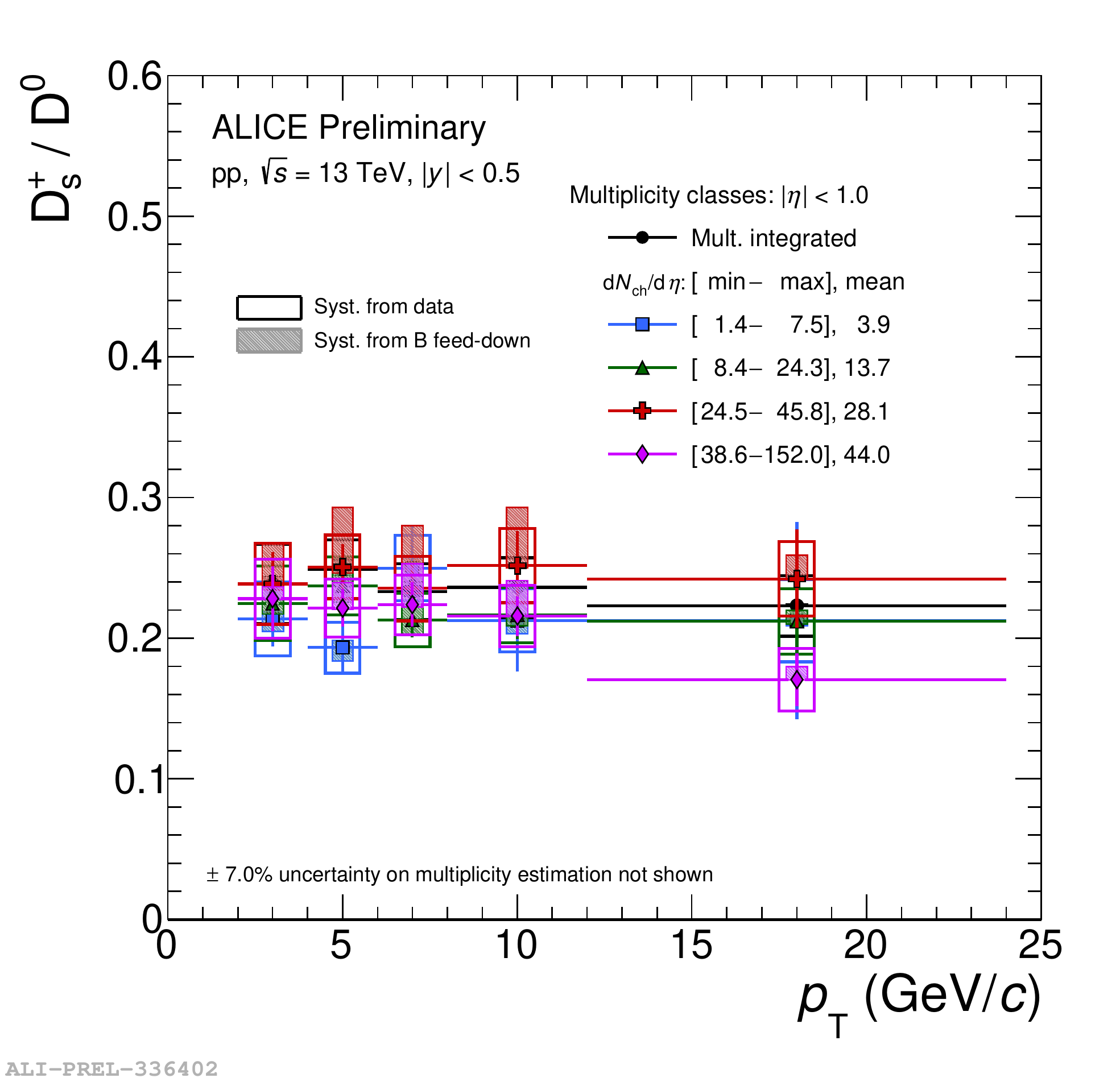}
		\includegraphics[width=.495\textwidth]{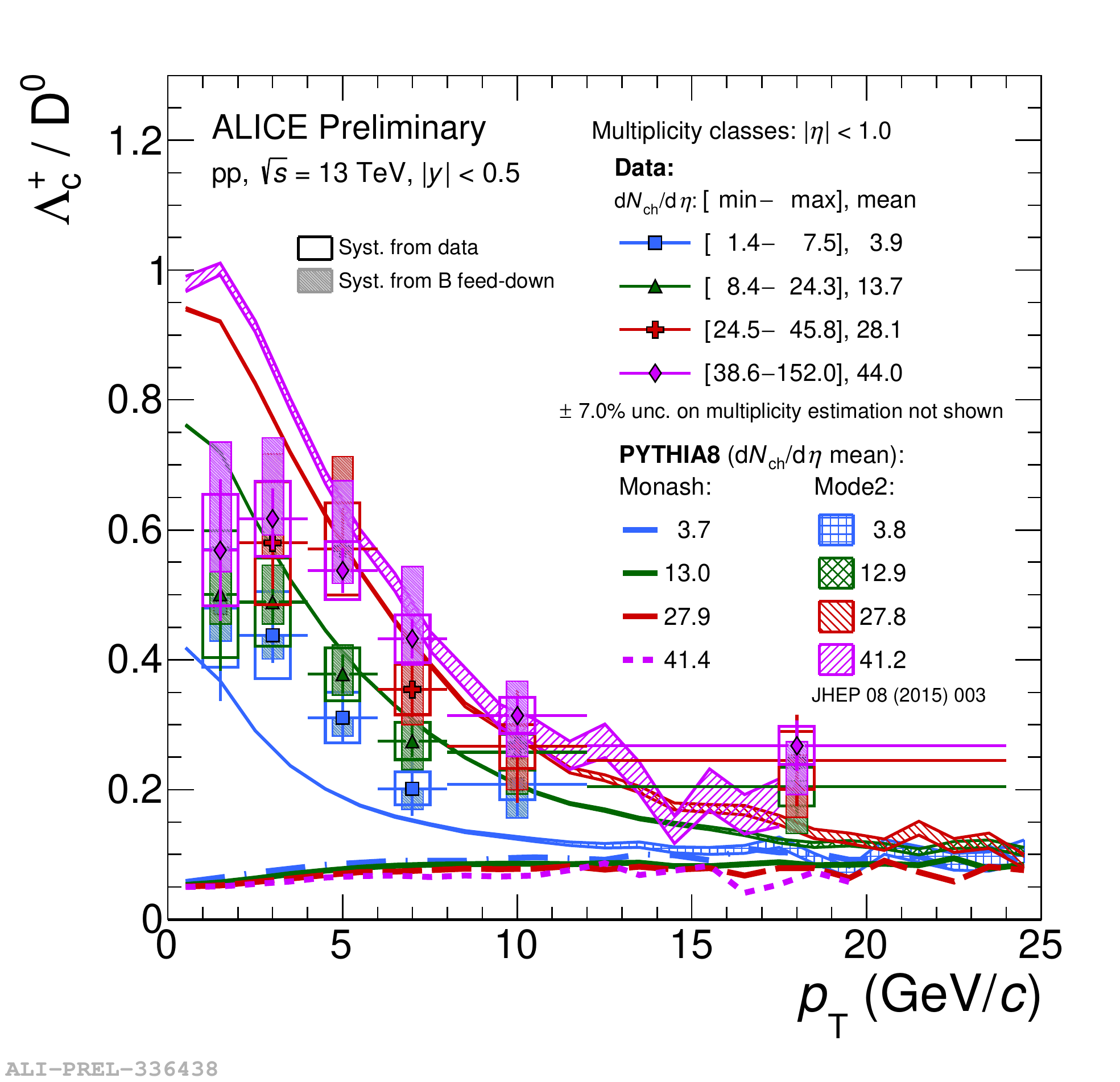}
		\caption{The ${\rm D}_{\rm s}^{+} / {\rm D}^{0}$ (left) and $\Lambda_{\rm c}^{+}/{\rm D}^{0}$ (right) ratios of prompt yields in pp collisions at $\sqrt{s}=13$~TeV measured in different multiplicity intervals. The $\Lambda_{\rm c}^{+}/{\rm D}^{0}$ ratios are compared to different \textsc{PYTHIA} calculations.}
		\label{fig:ppvsmult}
	\end{center}
\end{figure}

The production of charmed hadrons is measured in pp collisions at $\sqrt{s}=13$~TeV in different multiplicity intervals to study heavy-flavour hadronisation in small systems. Especially the charmed-hadron ratios ${\rm D}_{\rm s}^{+} / {\rm D}^{0}$ and $\Lambda_{\rm c}^{+}/{\rm D}^{0}$, shown in Fig.~\ref{fig:ppvsmult} as function of transverse momentum $p_{\rm T}$, can help investigating if recombination processes start to play already a role in high multiplicity pp collisions. Furthermore, low multiplicity pp collisions might shed light on the puzzle that is observed when comparing the $\Lambda_{\rm c}^{+}/{\rm D}^{0}$ measurements in minimum-bias pp and $\rm e^+e^-$ collisions~\cite{Acharya:2017kfy}. This ratio is found to be significantly enhanced in pp collisions compared to $\rm e^+e^-$ collision data, motivating studies of ultra low multiplicity pp events. In the presented measurements, the multiplicity is measured at midrapidity according to the number of tracklets reconstructed in the two innermost layers of the Inner Tracking System~\cite{Aamodt:2008zz}. The measured ${\rm D}_{\rm s}^{+} / {\rm D}^{0}$ ratios are consistent among each other within the experimental uncertainties, and are in agreement with the value expected considering charm-quark fragmentation fractions from ${\rm e^+ e^-}$ collisions~\cite{Gladilin:2014tba}. The $\Lambda_{\rm c}^{+}/{\rm D}^{0}$ ratios indicate a stronger multiplicity dependence. An enhancement compared to \textsc{PYTHIA} calculations with the Monash tune, which describe the $\Lambda_{\rm c}^{+}/{\rm D}^{0}$ measurements in $\rm e^+ e^-$ collisions, is still observed for the lowest multiplicity interval. It is interesting to see that the \textsc{PYTHIA} calculations with Mode2, which include colour reconnection beyond leading colour~\cite{Christiansen:2015yqa}, describe reasonably well the multiplicity enhancement. However, to draw a firm conclusion, higher precision measurements along with further theoretical comparisons are needed.

\begin{figure}[!tb]
	\begin{center}
		\includegraphics[width=0.5\textwidth]{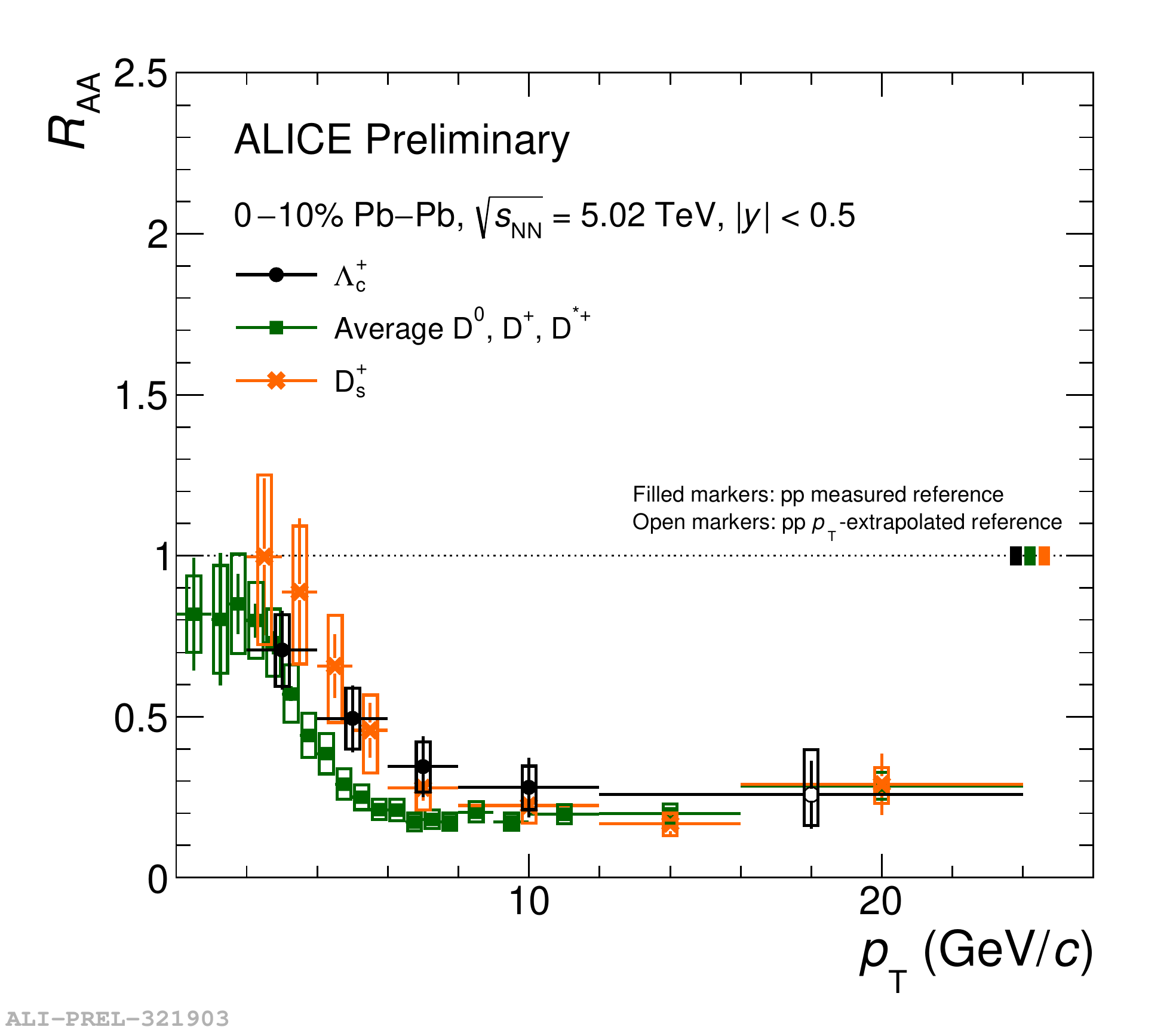}
		\includegraphics[width=0.475\textwidth]{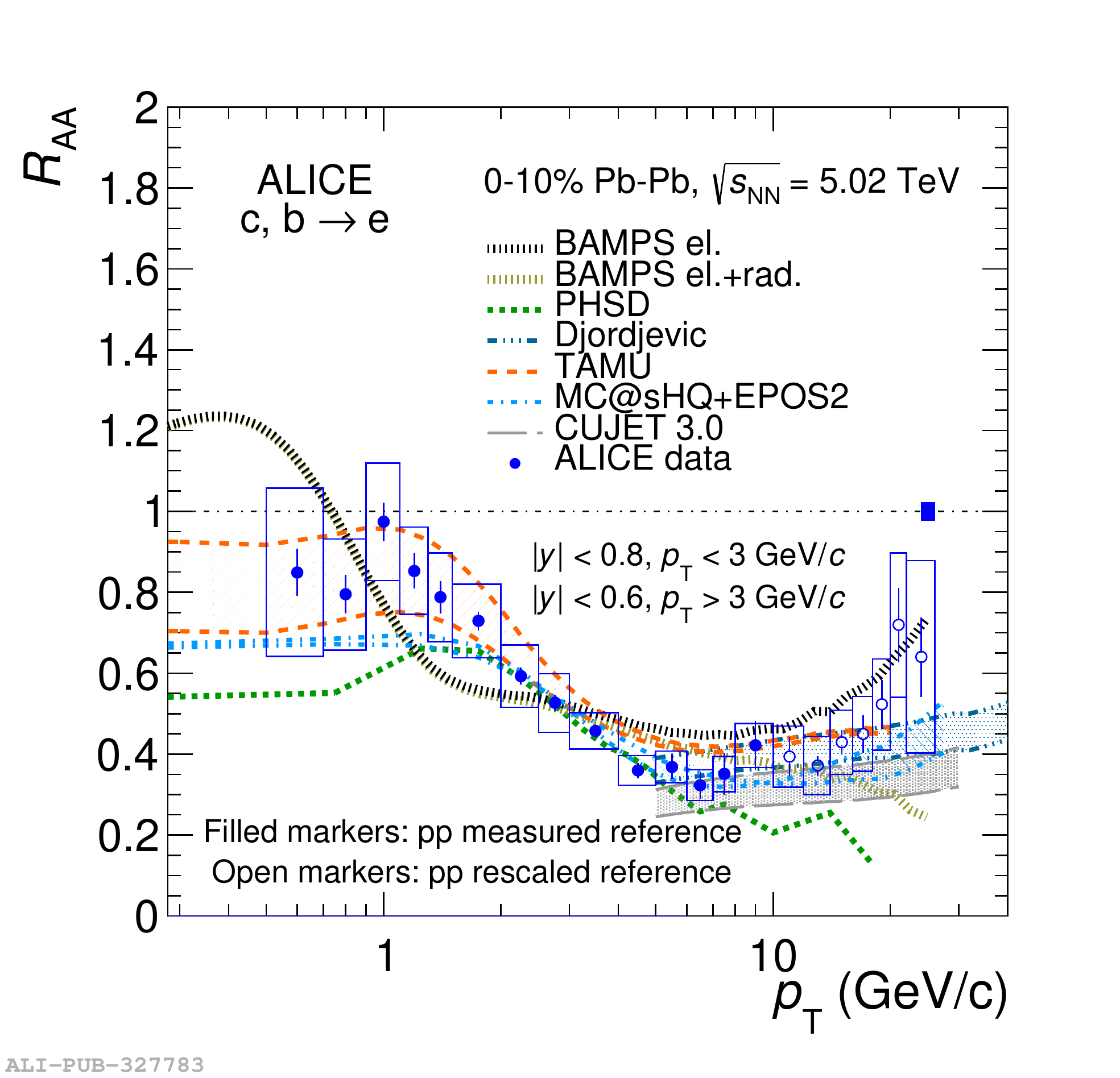}
		\caption{Left: Nuclear modification factor of  $\Lambda_{\rm c}^{+}$ baryons, non-strange ${\rm D}$ mesons, and ${\rm D}_{\rm s}^{+}$ mesons in the 0--10\% centrality class for Pb--Pb collisions at $\sqrt{s_{\rm NN}} = 5.02$~TeV. Right: Nuclear modification factor of $\rm e^{\pm}$ from heavy-flavour hadron decays compared to theoretical predictions~\cite{Acharya:2019mom}.}
		\label{fig:RaaHF}
	\end{center}
\end{figure}

The production of prompt charmed hadrons is also measured in Pb--Pb collisions at $\sqrt{s_{\rm NN}} = 5.02$~TeV, as well as the production of electrons from heavy-flavour hadron decays~\cite{Acharya:2019mom}. Figure~\ref{fig:RaaHF} shows the measured nuclear modification factor ($R_{\rm AA}$) in the 10\% most central Pb--Pb collisions compared between charmed hadrons (left panel) and theoretical predictions (right panel). The $R_{\rm AA}$ is defined as the ratio between the $p_{\rm T}$-differential yield measured in nucleus-nucleus collisions (${\rm d}^{2}N_{\rm AA}/{\rm d}p_{\rm T}{\rm d}y$) and the $p_{\rm T}$-differential production cross section in pp collisions (${\rm d}^{2}\sigma_{\rm pp}/{\rm d}p_{\rm T}{\rm d}y$), scaled by the average nuclear overlap function $\langle T_{\rm AA} \kern-0.1em\rangle$, a quantity proportional to the number of binary nucleon-nucleon collisions. It is used to study the properties of the in-medium parton energy loss. Strong suppressions of the $R_{\rm AA}$ are observed in Fig.~\ref{fig:RaaHF}, as expected in the presence of the QGP. There is an indication of less nuclear modification for ${\rm D}_{\rm s}^{+}$ mesons and $\Lambda_{\rm c}^{+}$ baryons compared to non-strange ${\rm D}$ mesons, although the uncertainties are still relatively large. The $R_{\rm AA}$ of electrons from heavy-flavour hadron decays is described by several transport models, indicating an interplay between a hydrodynamic medium expansion, collisional and radiative energy loss mechanisms, and hadronisation via recombination~\cite{Acharya:2019mom}.

\begin{figure}[!tb]
	\begin{center}
		\includegraphics[width=.47\textwidth]{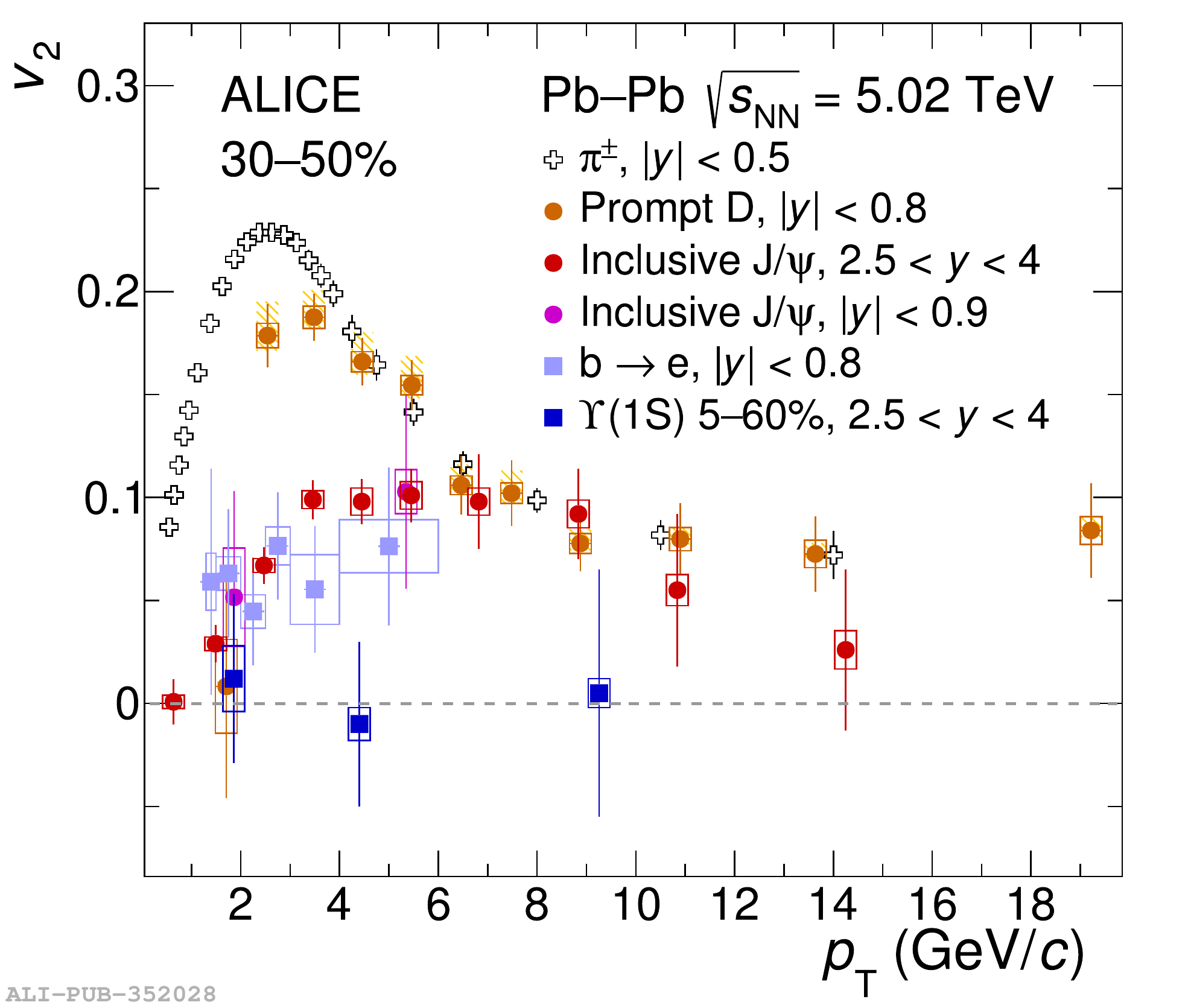}
		\hspace{0.02\textwidth}
		\includegraphics[width=.425\textwidth]{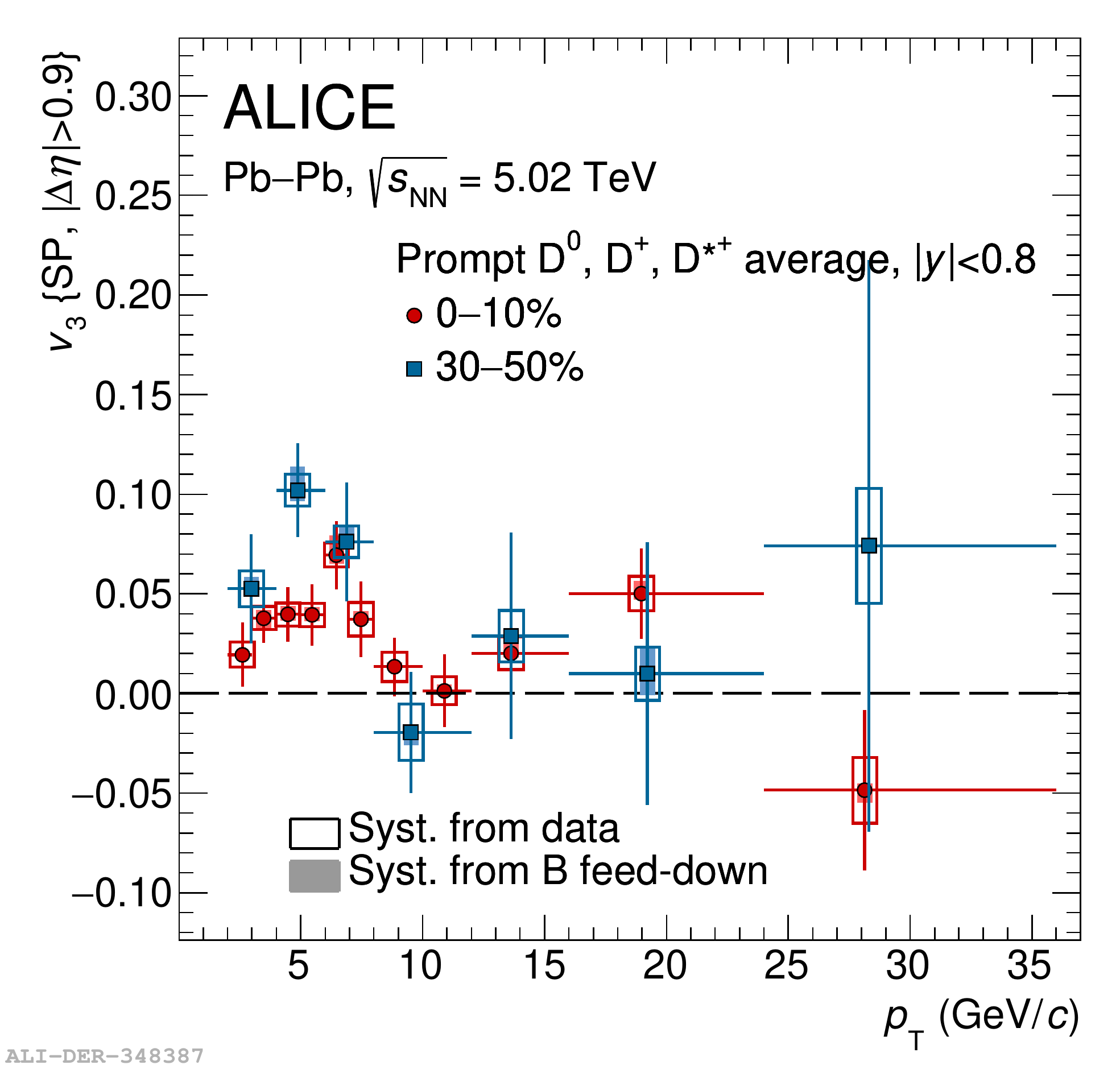}
		\caption{Left: The $v_{2}$ for charged pions, non-strange ${\rm D}$ mesons, inclusive $\rm J/\psi$, electrons from beauty-hadron decays, and $\Upsilon(\rm 1S)$ in 30--50\% central Pb--Pb collisions at $\sqrt{s_{\rm NN}} = 5.02$~TeV~\cite{Acharya:2020pnh,Acharya:2020qvg,Acharya:2020jil,Acharya:2019hlv}.  Right: The $v_3$ for non-strange ${\rm D}$ mesons in central (0--10\%) and semi-central (30--50\%) Pb--Pb collisions~\cite{Acharya:2020pnh}.}
		\label{fig:flowHF}
	\end{center}
\end{figure}

The measurements of azimuthal anisotropies, characterised by the magnitude of the Fourier coefficients $v_{\rm n} = \langle \cos{n (\varphi - \Psi_{\rm n})} \kern-0.1em\rangle$, can give further insights into the interaction of heavy quarks with the QGP. Here, $\varphi$ is the particle-momentum azimuthal angle, the brackets denote the average over all measured particles, and $\Psi_{\rm n}$ is the symmetry-plane angle relative to the n$^{\rm th}$ harmonic. The $v_2$, called elliptic flow, is the dominant harmonic coefficient in semi-central heavy-ion collisions and shown in the left panel of Fig.~\ref{fig:flowHF} for several particle species~\cite{Acharya:2020pnh,Acharya:2020qvg,Acharya:2020jil,Acharya:2019hlv}. The elliptic flow is sensitive to the degree of thermalisation of the heavy quarks at low $p_{\rm T}$, where at high $p_{\rm  T} $ it provides additional information on the path-length dependence of the in-medium parton energy loss. For $p_{\rm  T} < 3$~GeV$/c$, a mass ordering can be noticed, while for $3 < p_{\rm T} < 6$~GeV$/c$, the $v_{2}$ is similar for prompt ${\rm D}$ mesons and charged particles, indicating recombination of charm quarks with flowing light-flavour quarks. The measured $v_2$ coefficients seem to converge for all particle species for $p_{\rm T} > 6$~GeV$/c$, hinting to a similar path-length dependence for heavy and light parton energy loss. The triangular flow ($v_3$) is found to be positive for non-strange ${\rm D}$ mesons as shown in the right panel of Fig.~\ref{fig:flowHF}. It originates from event-by-event fluctuations in the initial distributions of participant nucleons and is more sensitive to the shear viscosity over entropy density ratio $\eta/s$ than the $v_2$. Transport models including both fragmentation and recombination fairly describe the $v_2$ and $v_3$ of ${\rm D}$ mesons, suggesting a range for the charm thermalisation time between 3 and 14 ${\rm fm}/c$~\cite{Acharya:2020pnh}. 

\begin{figure}[!tb]
	\begin{center}
		\includegraphics[width=.49\textwidth]{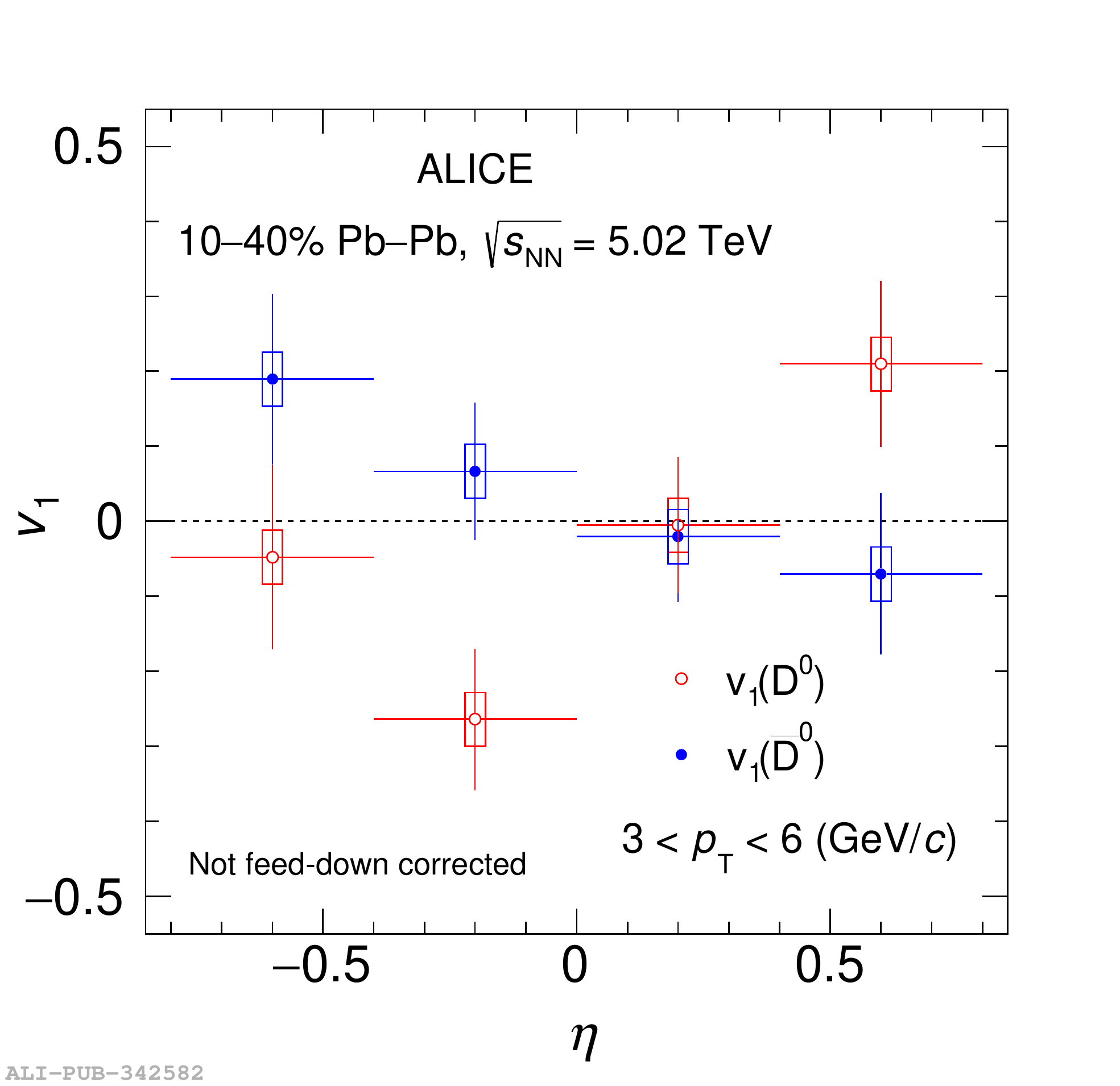}
		\includegraphics[width=.49\textwidth]{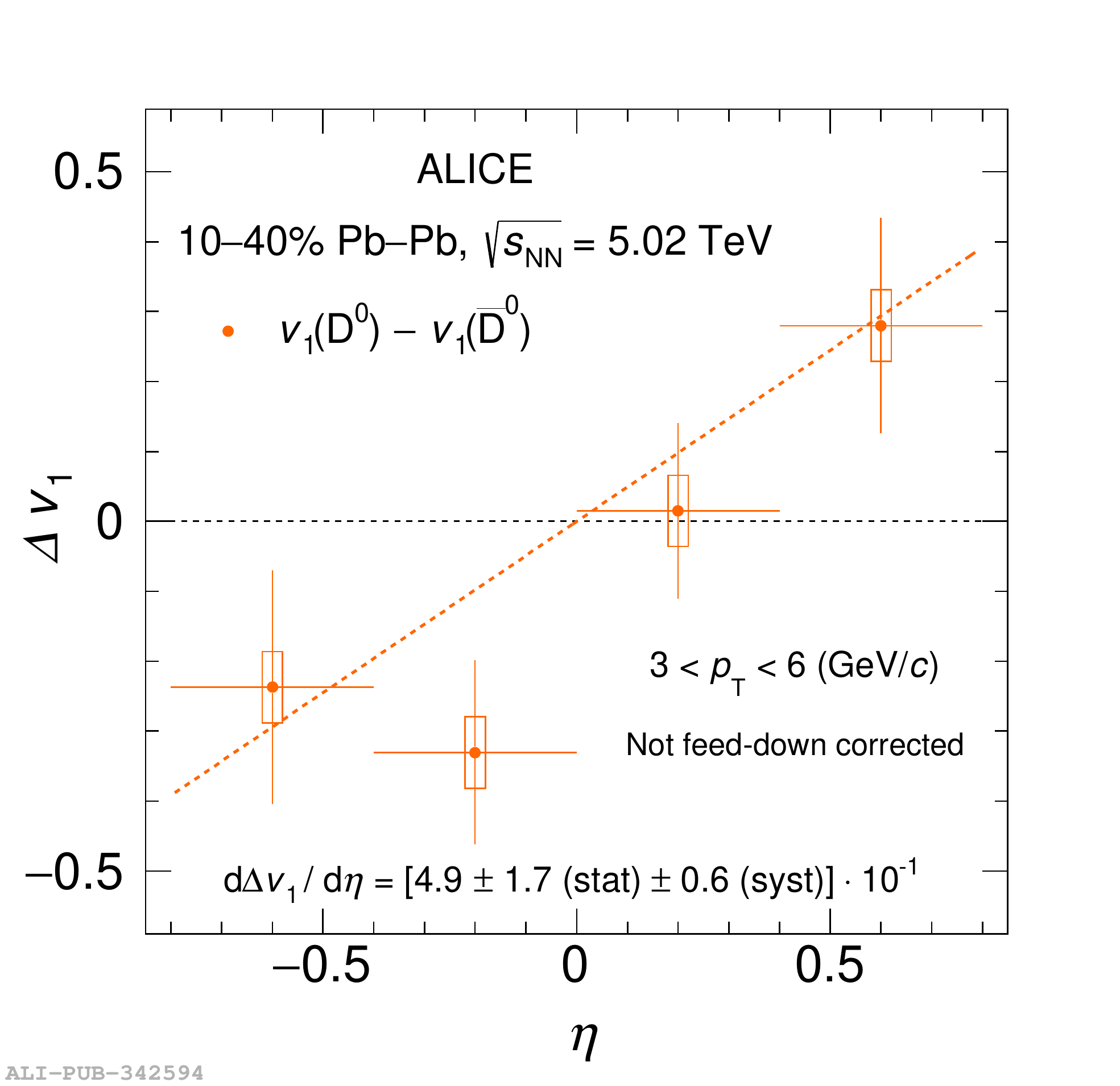}
		\caption{The directed flow for prompt ${\rm D^{0}}$ and $\rm \overline{D}$$^0$ mesons (left) and their difference $\Delta v_{1}$ (right) for $3 < p_{\rm T} < 6$~GeV$/c$ as a function of pseudorapidity in 10--40\% central Pb--Pb collisions~\cite{Acharya:2019ijj}.}
		\label{fig:v1D0}
	\end{center}
\end{figure}

The first harmonic coefficient ($v_1$), called directed flow, is measured for prompt ${\rm D^{0}}$ and $\rm \overline{D}$$^0$ mesons by the ALICE Collaboration as well~\cite{Acharya:2019ijj}. The directed flow of charmed hadrons is suggested as a powerful probe to study the strong electromagnetic fields produced in heavy-ion collisions due to the early production time of charm quarks and the charm thermalisation time, which is similar to the QGP lifetime. As a consequence of these fields, a charge-dependent contribution to the pseudorapidity-odd component of $v_1$ is expected~\cite{Dubla:2020bdz}. When measured, the charge-dependent directed flow may also provide important constraints to the QGP electric conductivity~\cite{Tuchin:2013ie}. The directed flow for prompt ${\rm D^{0}}$ and $\rm \overline{D}$$^0$ mesons, and their difference, is shown in Fig.~\ref{fig:v1D0} for 10--40\% central Pb--Pb collisions at $\sqrt{s_{\rm NN}} = 5.02$~TeV. The difference in $v_1$ is found to have a positive slope as a function of pseudorapidity $\eta$, which is larger than zero with a significance of 2.7$\sigma$ and about three orders of magnitude larger than the measured slope for charged hadrons~\cite{Acharya:2019ijj}. These measurements challenge some recent theoretical predictions that predicted an order of magnitude smaller and negative values of ${\rm d}\Delta v_1/{\rm d}\eta$~\cite{Das:2016cwd}.

\section{Conclusions}
In this contribution, the most recent results on open heavy-flavour production in pp and Pb--Pb collisions measured by the ALICE Collaboration were presented. These new measurements are performed on Run 2 data samples, and are measured for the first time or with an improved statistical precision and more differential with respect to earlier ALICE publications. The results show that hadronisation of charm quarks is already modified in pp collisions, that charm quarks experience strong nuclear modification in the QGP, that they participate in the collective motion of the system, and are sensitive to the strong magnetic fields present during the collision. Furthermore, hadronisation via recombination seems to be necessary in the theoretical models to quantitatively reproduce charmed-hadron production in Pb--Pb collisions.

\end{document}